\documentclass[aps,pra,twocolumn]{revtex4-1}
\usepackage{newlfont}
\usepackage{amssymb}
\usepackage{amsfonts}
\usepackage{amsmath}
\usepackage{amsthm}
\usepackage{graphicx}
\usepackage{epsfig}
\usepackage{color}
\usepackage{bm}

\begin{document}
\title{Light Cone-Like Behavior of Quantum Monogamy Score and Multisite Entanglement}

\author{R. Prabhu, Arun Kumar Pati, Aditi Sen(De), and Ujjwal Sen}

\affiliation{Harish-Chandra Research Institute, Chhatnag Road, Jhunsi, Allahabad 211 019, India}


\begin{abstract}

There are two important paradigms for defining quantum correlations in quantum information theory, viz. the 
information-theoretic and the entanglement-separability ones. 
We find an analytical relation between two measures of quantum correlations, one in each paradigm, and show that only a certain cone-like region on the two-dimensional space spanned by these 
measures is accessible to pure three-qubit states. 
The information-theoretic multiparty quantum correlation measure is related to the monogamy considerations of a bipartite information-theoretic quantum correlation measure, while the entanglement-separability multiparty measure is 
the generalized geometric measure, a genuine multiparty entanglement measure.
We also find an analytical relation between two multiparty entanglement measures, and again obtain a cone-like accessible region in this case. 
One of the multisite measures in this case is related to the monogamy of a bipartite entanglement measure, while the other is again the generalized geometric measure. 
Just like in relativity, events cannot occur outside the space-time light cone, we analogously find here that state points corresponding to pure three-qubit states cannot fall outside the two-dimensional cone-like structure between 
quantum monogamy scores and a genuine multisite entanglement measure.

\end{abstract}

\maketitle

\section{Introduction and Main Results}
In the past few decades, there have been several discoveries in the field of quantum information science 
which utilize quantum mechanical principles to enhance 
our abilities to compute and communicate  \cite{ekhane-NC}. 
%
An important connecting string in this area is the notion of 
``quantum correlation'' that may exist in the multiparticle quantum states.
 In particular, entanglement of shared quantum states \cite{HHHH-RMP} is the vital element for the success of quantum communication protocols \cite{comm-review}, 
including the quantum dense coding protocol \cite{dc}, quantum teleportation \cite{tele}, entanglement-based quantum cryptography \cite{Ekert91}, and
remote state preparation \cite{remote}. Shared entanglement is also an essential ingredient of measurement-based quantum computation \cite{Briegel}. 
The theoretical success of quantum information protocols has been closely followed by experimental implementations, and in particular, 
entangled multisite quantum states are being realized in a large number of laboratories.

While most multiparty quantum information phenomena occur with the active support of shared entanglement, there are important examples where a multisite nonclassical 
phenomena is predicted without using entangled quantum states. One example is the phenomenon of ``quantum nonlocality without entanglement'', where a set of unentangled 
multiparty quantum states exhibit the nonclassical feature of local indistinguishability of orthogonal states \cite{nlwe} (cf. \cite{local-indis-other}). 
Another example is  deterministic quantum computation with one quantum bit, where it is possible to perform simulations that have no known efficient classical algorithms, 
although the control qubit and the mixed qubits do not possess any entanglement between themselves \cite{KnillLaflamme,Animesh,others}. 
It is therefore important to search for quantum correlation concepts that are independent of the entanglement-separability paradigm. Such attempts have been made, a 
prominent one being quantum discord, defined from information-theoretic concepts \cite{discord1, discord2}. 
There have been several other developments in this direction including quantifying multiparticle quantum correlations \cite{workdeficit, Modi, Arun-disso, discord-multi, others, bakisob}.

Quantum correlations defined within the entanglement-separability paradigm and from an information-theoretic perspective are apparently very different in nature, 
and exhibit quite a variety of different mutually exclusive properties. It is therefore interesting to find common features shared by all or a class of quantum correlation measures. 
In this direction, it may be envisaged that quantum correlations, in contrast to classical ones, satisfy a monogamy relation. Such a relation is indeed satisfied by 
certain measures of entanglement, so that if two parties are highly entangled, then they cannot have a large amount of entanglement shared with a third one 
\cite{Wootters, Koashi, monogamyN}. 
However, monogamy may not be satisfied by information-theoretic measures of quantum correlations \cite{monogamy-discord} (cf. \cite{monogamy-iacc}). 
More precisely, the monogamy relation for a bipartite quantum correlation measure \({\cal Q}\), as applied to a quantum state \(\rho_{ABC}\) shared between three 
observers \(A\), \(B\), and \(C\), states that 
\begin{equation}
{\cal Q}(\rho_{AB}) + {\cal Q}(\rho_{AC}) \leq {\cal Q}(\rho_{A:BC}),
\end{equation}
where \(\rho_{AB} = \mbox{tr}_C \rho_{ABC}\) and similarly for \(\rho_{AC}\), and \({\cal Q}(\rho_{A:BC})\) is the measure \({\cal Q}\) of the state \(\rho_{ABC}\) 
considered in 
the \(A:BC\) bipartite split. We introduce here the concept of ``quantum monogamy score'', corresponding to a quantum correlation measure \({\cal Q}\), given by 
\begin{equation}
\delta_{{\cal Q}} = {\cal Q}(\rho_{A:BC}) - {\cal Q}(\rho_{AB}) - {\cal Q}(\rho_{AC}), 
\end{equation}
so that the tripartite quantum state \(\rho_{ABC}\) satisfies the monogamy relation for \({\cal Q}\) if the quantum monogamy score, corresponding to \({\cal Q}\), 
is positive, and violates the same if \(\delta_{{\cal Q}}\) is negative. Note here that the definition of the monogamy relation and quantum monogamy 
score gives a special status to one of the three observers (observer \(A\), here). 
We will call such an observer as the ``node'' for the particular quantum monogamy score defined.
For some quantum correlation measures, the quantum monogamy score may be independent of which observer  is considered to be the nodal observer. However, this is 
not true in general.


In an effort towards finding quantitative connections between the twin paradigms, entanglement-separability and information-theoretic,
 in which quantum correlations are defined,
Koashi and Winter \cite{Koashi} established a relation between a \emph{bipartite} entanglement measure (precisely, entanglement of formation \cite{Eof, concurrence})
  and  a \emph{bipartite} information-theoretic quantum correlation measure (precisely, quantum discord) for three-qubit states. 
A similar connection was obtained in Ref. \cite{discord-surge} between logarithmic negativity \cite{logneg} (a bipartite entanglement measure) and bipartite 
information-theoretic
quantum correlation measures, in the quantum dynamics of a spin chain. 
We address the same question in a \emph{multipartite} scenario. In particular, we obtain a relation between a multipartite 
information-theoretic quantum correlation (``discord monogamy score'') and a genuine multipartite entanglement 
measure (generalized geometric measure (GGM) \cite{GGM}) for three-qubit pure states.
We reveal a cone-like structure in the space spanned by the two measures. 
More precisely, given a certain amount of quantum monogamy 
score, we find that the genuine multiparty entanglement (as quantified by GGM) of a three-qubit pure quantum state \(|\psi\rangle\) is restricted to lie above 
a certain (nonzero) positive value. Interestingly, this positive value coincides with the 
GGM of the generalized GHZ state \cite{GHZ}, whose quantum monogamy score is equal to the modulus of that of \(|\psi\rangle\). 
The quantum correlation measure in the quantum monogamy score is quantified either by the square of the 
concurrence \cite{concurrence} or by quantum discord \cite{discord1, discord2}. 

Our analysis shows that in analogy with the space-time light cone where events cannot occur outside it, 
multiparty quantum states cannot appear outside the ``light cone'' of quantum monogamy scores and genuine multiparty entanglement.
%
%
We believe that such relations will help towards building a unified framework for quantum correlation measures.
That the quantum monogamy score for concurrence squared is a multiparty entanglement measure was already noted by Coffman, Kundu, and Wootters \cite{Wootters}. Below we will show (Proposition I) that 
the quantum monogamy score corresponding to quantum discord can be interpreted as an information-theoretic multiparty quantum correlation measure. 

We begin the next section (Sec. \ref{sec-amra-Mukunda-r-ghore-bosey-achhi-Bengaluru-te}) by providing brief sketches of the three 
measures of quantum correlations, that will be required for the rest of the paper. 
The results are presented in 
Sec. \ref{sec-Pasupathy-Harappa-thheke-IISc}, where we find the relations between
quantum monogamy score (for concurrence squared as well as for quantum discord) and a genuine multiparty entanglement for 
pure three-qubit states. In Sec. \ref{poune-baro-at-4-Ishan-Mondol-1}, we establish our results analytically. Numerical simulations are presented in 
Secs. \ref{poune-baro-at-4-Ishan-Mondol-2} and  \ref{poune-baro-at-4-Ishan-Mondol-3}. Two plots are generated: 
The  plot between quantum monogamy score for concurrence squared and 
GGM
for randomly generated three-qubit pure states is discussed in  Sec. \ref{poune-baro-at-4-Ishan-Mondol-2}, while that  
between quantum monogamy score for quantum discord and
GGM for the same states is discussed in Sec. \ref{poune-baro-at-4-Ishan-Mondol-3}.
In Sec. \ref{holde-sobuj-orangotang}, we show that the quantum monogamy score for quantum discord can be seen as a multisite quantum correlation measure, defined from an information-theoretic perspective.
The analytical relation connecting the quantum monogamy score for concurrence squared (called ``entanglement monogamy score'' herein; had also been named 3-tangle) 
and the generalized geometric measure is presented in Sec. \ref{lal-akash}, while that connecting the 
quantum monogamy score for quantum discord (called ``discord monogamy score'' herein) and the generalized geometric measure is given at Eq. (\ref{sobuj-prithhibi}) of Sec. \ref{holde-sobuj-orangotang}.
We discuss our results in a concluding section (Sec. \ref{sec-ekhane-conclu}).

\section{Measures of Quantum Correlation}
\label{sec-amra-Mukunda-r-ghore-bosey-achhi-Bengaluru-te}

In this section, we will briefly describe the measures of quantum correlations that will be used later in this paper.

\subsection{Concurrence}

The concept of concurrence \cite{Wootters,concurrence} originates from the definition of entanglement of formation.
The entanglement of formation of a bipartite quantum state is intuitively (modulo certain 
additivity problems) the amount of singlets, \(\frac{1}{\sqrt{2}}(|01\rangle - |10\rangle)\),  that are required to prepare 
the state by local quantum operations and classical communication (LOCC). Here, \(|0\rangle\) and \(|1\rangle\) are 
orthonormal quantum states.
The entanglement of formation of a pure bipartite state, \(|\varphi\rangle_{AB}\) shared between two parties \(A\) and \(B\),
 can be shown to be equal to the von Neumann entropy of 
the local density matrix of the shared state \cite{Eof}: 
\begin{equation}
E(|\varphi\rangle_{AB})= S(\varrho_A) = S(\varrho_B).
\end{equation} 
Here $\varrho_{A}$ and \(\varrho_B\) are the partial traces of combined system $|\psi\rangle_{AB}$ over subsystems $B$ and $A$ respectively, and 
\(S(\sigma) = - \mbox{tr} \left(\sigma \log_2 \sigma\right)\) is the von Neumann entropy of a quantum state \(\sigma\). 
Entanglement of formation of a mixed bipartite state \(\rho_{AB}\) is 
then defined by the convex roof approach: 
\begin{equation}
E(\rho)=\mbox{min}\sum_i p_iE(|\varphi_i\rangle),
\end{equation}
where the minimization is over all pure state decompositions of $\rho = \sum_i p_i (|\varphi_i\rangle \langle \varphi_i|)_{AB}$.

This minimization is usually hard to perform. However, there exists a closed form in the case of two-qubit states \cite{concurrence}, in terms of 
the concurrence.  
The concurrence $C(\rho)$ is defined as $C(\rho)=\mbox{max}\{0,\lambda_1-\lambda_2-\lambda_3-\lambda_4\}$, where the
$\lambda_i$'s are the square roots of the eigenvalues of $\rho\tilde{\rho}$ in decreasing order. Here \(\tilde{\rho}\) is given by 
$\tilde{\rho}=(\sigma_y\otimes\sigma_y)\rho^*(\sigma_y\otimes\sigma_y)$, where the complex conjugation is performed
in the computational  basis and $\sigma_y$ is the Pauli spin matrix.

\subsection{Quantum Discord}

Classically, there are two equivalent ways to arrive at the concept of mutual information between two random variables. One is by adding the 
Shannon entropies of the individual 
random variables, and then subtracting that of the joint probability distribution. Therefore, 
%
the 
mutual information \(H(X:Y)\) between two random variables \(X\) and \(Y\) can be defined as 
\begin{equation}
\label{mutualinfo1}
H(X:Y) = H(X) + H(Y) - H(X,Y).
\end{equation}
Here \(H(X) = -\sum p_i \log_2 p_i\) is the Shannon entropy of the random variable \(X\) that is distributed according to the 
probability distribution \(\{p_i\}\), and
\(H(X,Y)\) is the Shannon entropy of the joint probability distribution of the two random variables \(X\) and \(Y\).

A second way is to interpret the Shannon entropy of a random variable as the information deficit (disorder) that we possess for that random variable. 
Then we choose any of the two random variables, say 
\(X\), and consider the information deficit, \(H(X)\), corresponding to that random variable. Then, the 
 mutual information between \(X\) and \(Y\) can be defined as the 
disorder remaining in the variable \(X\) after the ``conditional disorder'' of the variable \(X\) given that \(Y\) has already occurred, is removed. 
Precisely, this is given by 
%
%
\begin{equation}
\label{mutualinfo2}
H(X:Y) = H(X) - H(X|Y),
\end{equation}
where \(H(X|Y)\) 
is the corresponding conditional entropy. 
Of course, these two ways of defining the (classical) mutual information are mathematically equivalent.

In the quantum domain, these two quantities are different, and leads to the definition of quantum discord
\cite{discord1, discord2}. The classical definition of mutual information given in Eq. (\ref{mutualinfo1}) can be 
taken over to the quantum regime 
by replacing the Shannon entropies by von Neumann ones: For a quantum state \(\rho_{AB}\) of two parties, the ``quantum mutual information'' 
is defined as \cite{qmi} (see also \cite{Cerf, GROIS})
\begin{equation}
\label{qmi1}
I(\rho_{AB})= S(\rho_A)+ S(\rho_B)- S(\rho_{AB}),
\end{equation}
where \(\rho_A\) and \(\rho_B\) are the local density matrices of \(\rho_{AB}\).

Quantization of the second definition of classical mutual information (Eq. (\ref{mutualinfo2})) has to be performed in a different way than 
replacing Shannon entropies by von Neumann ones, as the latter gives rise to
a physical quantity that can be  
 negative for some quantum states \cite{Cerf}.
However, interpreting the conditional entropy in the classical case, as a measure of the lack of information about one random variable 
when the other is known, in a joint probability distribution of two random variables, 
the second definition of classical mutual information can be quantized for 
a bipartite quantum state \(\rho_{AB}\)
as
\begin{equation}
\label{cmi1}
J(\rho_{AB}) = S(\rho_A) - S(\rho_{A|B}),
\end{equation}
where the ``quantum conditional entropy'' is defined as  
\begin{equation}
\label{qce}
S(\rho_{A|B}) = \min_{\{\Pi_i^B\}} \sum_i p_i S(\rho_{A|i}),
\end{equation}
with the minimization being over all 
projection-valued measurements, \(\{\Pi^B_i\}\),  performed on subsystem \(B\).
Here \(p_i = \mbox{tr}_{AB}(\mathbb{I}_A \otimes \Pi^B_i \rho_{AB} \mathbb{I}_A \otimes \Pi^B_i)\) is the probability for obtaining the outcome \(i\), and 
the corresponding post-measurement state  
for the subsystem \(A\) is \(\rho_{A|i} = \frac{1}{p_i} \mbox{tr}_B(\mathbb{I}_A \otimes \Pi^B_i \rho_{AB} \mathbb{I}_A \otimes \Pi^B_i)\), 
where \(\mathbb{I}_A\) is the identity operator on the Hilbert space of the quantum system that is in possession of \(A\).


It turns out that the two quantizations produce inequivalent quantum quantities, and is interpreted as the result of quantum correlations present in the bipartite quantum state. 
The difference was consequently interpreted as a measure of quantum correlations, and called as the quantum discord. 
Moreover, it was shown that the quantum mutual information is never lower than the quantity \(J\). Therefore, the 
quantum discord is given by 
\cite{discord1, discord2}
\begin{equation}
D(\rho_{AB})= I(\rho_{AB}) - J(\rho_{AB}).
\end{equation}
In contrast to many other measures of quantum correlations, even some separable states produce a nonzero discord.

\subsection{Generalized Geometric Measure}

A multiparty pure quantum state is said to be \emph{genuinely} multiparty entangled if it is entangled across every 
bipartition of its constituent parties. 
The amount of genuine multiparty entanglement present in a multiparty state 
can be quantified by the recently introduced genuine multipartite entanglement measure
called the generalized geometric measure (GGM) \cite{GGM} (cf. \cite{GM}).
The GGM of  an \(N\)-party pure quantum state \(|\phi_N\rangle\) is defined as
\begin{equation}
{\cal E} ( |\phi_N\rangle ) = 1 - \Lambda^2_{\max} (|\phi_N\rangle ), 
\end{equation}
where  \(\Lambda_{\max} (|\phi_N\rangle ) =
\max | \langle \chi|\phi_N\rangle |\), with  the maximization being over all pure states \(|\chi\rangle\)
that are not genuinely \(N\)-party entangled. 
It was shown in Ref. \cite{GGM} that 
\begin{equation}
\label{label}
{\cal E} (|\phi_N \rangle ) =  1 - \max \{\lambda^2_{{\cal A}: {\cal B}} |  {\cal A} \cup  {\cal B} = 
\{1,2,\ldots, N\},  {\cal A} \cap  {\cal B} = \emptyset\},
\end{equation}
where \(\lambda_{{\cal A}:{\cal B}}\) is  the maximal Schmidt coefficients in the \({\cal A}: {\cal B}\) 
bipartite split  of \(|\phi_N \rangle\).


\section{Quantum Monogamy Scores and GGM}
\label{sec-Pasupathy-Harappa-thheke-IISc}

We will consider the quantum monogamy scores for two measures of quantum correlations. 
The first one is when the measure of quantum correlation is the concurrence squared, 
while the second one is when it is the quantum discord. In the case when the quantum correlation measure is the square of the concurrence, 
the corresponding quantum monogamy score has been called the ``3-tangle'' or ``residual tangle'' \cite{Wootters, concurrence}. We will call it the ``entanglement monogamy score'', 
and denote it as \(\delta_{C}\). 
In the second case, when the quantum correlation measure is the quantum discord, the quantum monogamy score is related to the concept of  ``dissension'' \cite{Arun-disso}, 
a multiparty information-theoretic quantum correlation measure. We will explicitly show the connection between the quantum monogamy score 
for quantum discord with dissension-like multiparty quantum correlation measures in Sec. \ref{holde-sobuj-orangotang}.
 We will call the quantum monogamy score corresponding to quantum discord as the ``discord monogamy score'', and denote it as \(\delta_{D}\). 

Intuitively, the quantum monogamy score, corresponding to a chosen 
bipartite quantum correlation measure, of a multiparty quantum system, 
encapsulates a \emph{multiparty}
quantum correlation of the shared system. For example for a tripartite 
quantum state shared between \(A\), \(B\), and \(C\), the 
quantum monogamy score is the amount of quantum correlation remaining 
in the \(A:BC\) bipartite split, after the two bipartite 
contributions for \(A:B\) and \(A:C\) are subtracted out. 
The remaining quantum correlations must therefore be multipartite in 
nature.  

It is therefore natural to look for connections of quantum monogamy score
with other more directly-defined measures of multiparty quantum correlations, 
like the generalized geometric measure. It is interesting to use 
the generalized geometric measure as a measure of multiparty quantum correlations
because 
\begin{itemize}

\item[(a)]  it is a measure of \emph{genuine} multiparty entanglement, and 

\item[(b)] it is easy to compute for pure states of an arbitrary number of 
parties and in arbitrary dimensions.
\end{itemize} 

\subsection{Relation between quantum monogamy scores and genuine multipartite entanglement measure}
\label{poune-baro-at-4-Ishan-Mondol-1}

In this subsection, we will establish a structure in the space spanned by the quantum monogamy score for a chosen quantum correlation measure and GGM for 
arbitrary three-qubit pure states. We will show that arbitrary points in these two-dimensional spaces are not accessible to quantum states. Moreover, it turns out 
that the generalized GHZ states form the boundaries of the respective accessible regions. The class of generalized GHZ states is formed by the states 
\begin{equation}|GG(\alpha)\rangle = \alpha |000\rangle +\beta |111\rangle,
 \end{equation}
 with $\alpha$ and \(\beta\) being real and positive, and $\alpha^2+\beta^2=1$ \cite{GHZ}.
Without loss of generality, we assume that \(\alpha \geq \beta\). 
Here \(|0\rangle\) and \(|1\rangle\) are two orthonormal states. We prove the analytical relations for the cases 
when the chosen quantum correlation measure in the definition of 
quantum monogamy score is either concurrence squared or quantum discord.

\subsubsection{Concurrence squared as the quantum correlation measure in quantum monogamy score}
\label{dos-ta-beyallis-atKGMarg}
We  introduce some notations. Consider an arbitrary pure three-qubit state \(|\psi_{ABC}\rangle\). We will henceforth drop the suffix, and denote it as \(|\psi\rangle\).
Let us denote its entanglement 
monogamy score as \(\delta_C\), remembering that it depends on the state \(|\psi\rangle\). Further, let us denote its GGM as ${\cal E}$, again remembering that 
it depends on the state \(|\psi\rangle\). Consider now the generalized GHZ state \(|GG(\alpha)\rangle\), and let us denote its entanglement monogamy score 
and GGM as $\delta_{C}^{GG}$ and 
${\cal E}^{GG}$  respectively, remembering that they depend on the generalized GHZ state parameter \(\alpha\).

For three-qubit pure states, the GGM of \(|\psi\rangle\) reduces to 
\begin{equation}
\label{ekhono-Mughalsarai-aseni}
 {\cal E} = 1-\max\{\lambda_A^2, \lambda_B^2, \lambda_C^2\},
\end{equation}
where \(\lambda_A^2\) is the maximal eigenvalue of \(\rho_A^\psi = \mbox{tr}_{BC}|\psi\rangle \langle \psi|\), and similarly for 
\(\lambda_B^2\) and \(\lambda_C^2\).
We will see below that ``the party (among \(A\), \(B\), and \(C\)) which contributes the maximal Schmidt coefficient in the GGM'',
that is the party (among \(A\), \(B\), and \(C\)) whose maximal eigenvalue of local density matrix attains the maximum in Eq. (\ref{ekhono-Mughalsarai-aseni}),
has an intimate connection with the 
quantum monogamy score.



We now prove that the entanglement monogamy score and GGM of arbitrary three-qubit pure states are constrained to lie within a cone-like structure as stated in the 
following theorem.\\
\noindent \textbf{Theorem 1:} Consider the pure three-qubit state \(|\psi\rangle\) whose entanglement monogamy score is the 
same as that of the generalized GHZ state \(|GG(\alpha)\rangle\). Then the genuine multipartite entanglement measures (GGM) of these two states will obey the ordering
\begin{equation}
{\cal E} \geq {\cal E}^{GG},
\end{equation}
independent of which observer is considered to be the 
nodal observer in the entanglement monogamy score.\\

\noindent \texttt{Proof:} 
The entanglement monogamy score for the three-qubit pure state \(|\psi\rangle\) is defined as
\begin{equation}
\delta_C = C^2_{A:BC} (|\psi\rangle) - C^2_{AB}(|\psi\rangle) - C^2_{AC}(|\psi\rangle), 
\end{equation}
 where \(A\) is considered to be the nodal observer.
Here, $C^2_{AB}(|\psi\rangle)$ and $ C^2_{AC} (|\psi\rangle)$ are the concurrence squared 
of the reduced density matrices  \(\rho_{AB}^\psi = \mbox{tr}_C |\psi\rangle \langle \psi|\) 
and \(\rho_{AC}^\psi= \mbox{tr}_B |\psi\rangle \langle \psi|\) 
of  \(|\psi\rangle\) respectively, and 
\(C^2_{A:BC} (|\psi\rangle)\) is the concurrence squared of \(|\psi\rangle\) in the \(A:BC\) split. 
One can show that  \cite{Wootters}  
\begin{equation}
C^2_{A:BC}(|\psi\rangle) = 4 \det \rho_A^\psi. 
\end{equation}
On the other hand, for the generalized
GHZ state \(|GG(\alpha)\rangle\), we have 
\begin{equation}
\delta_C^{GG}= 4 \det \rho_{A}^\alpha 
\end{equation}
 (with \(A\) being considered as the nodal observer), 
where 
\begin{equation}
\rho_A^\alpha = \mbox{tr}_{BC}|GG(\alpha)\rangle \langle GG(\alpha)|
= \alpha^2 |0\rangle \langle 0| + \beta^2 |1\rangle \langle 1|,       
\end{equation}
 so that 
\begin{equation}
 \delta_C^{GG} = 4\alpha^2(1-\alpha^2).
\end{equation} 
The enunciation is that 
\begin{equation}
\delta_C = \delta_C^{GG},                         
\end{equation}
which, for \(A\) as the nodal observer, implies that 
\begin{eqnarray}
\label{Nawapatra-ba-eirokom-kichhu-ekta-namer-station-gyalo}
\lambda_A^2 (1 - \lambda_A^2) \geq \alpha^2 (1 - \alpha^2).
\end{eqnarray}
To obtain the inequality, we have used the fact that the concurrences of $\rho_{AB}^\psi$ and $\rho_{AC}^\psi$ are 
nonnegative. Now $\delta_C$ is invariant under permutation of the parties \cite{Wootters},
using which we get two more similar inequalities pertaining 
to the nodal observers $B$ and $C$ respectively, and these inequalities are 
\begin{eqnarray}
\label{nodalBC}
\lambda_B^2 (1 - \lambda_B^2) &\geq& \alpha^2 (1 - \alpha^2),\nonumber\\
\lambda_C^2 (1 - \lambda_C^2) &\geq& \alpha^2 (1 - \alpha^2).
\end{eqnarray}
%
%
Let us now assume  that the GGM of the pure three-qubit state \(|\psi\rangle\) 
is strictly less than that of the generalized GHZ state \(|GG(\alpha)\rangle\), i.e., the proposed ordering is violated.
Now, suppose that the maximum in Eq. (\ref{ekhono-Mughalsarai-aseni}) is attained in \(\lambda_A^2\), i.e., \(\lambda_A^2 \geq \lambda_B^2, \lambda_C^2\).
Then, from the definition of GGM and some simple algebra, we get 
\begin{equation}
\lambda_A^2 (1 - \lambda_A^2) < \alpha^2 (1 - \alpha^2), 
\end{equation}
which is in contradiction with the inequality in (\ref{Nawapatra-ba-eirokom-kichhu-ekta-namer-station-gyalo}). 
Similar inequalities that are contradictory to the inequalities in  (\ref{nodalBC}) 
can be obtained when the maximum is attained by \(\lambda^2_B\) or \(\lambda_C^2\).
\hfill $\blacksquare$

\noindent \textbf{Remark:} The entanglement monogamy score of an arbitrary pure three-qubit state lies between 0 and 1. And for all \(\epsilon \in [0,1]\), 
there is a generalized GHZ state, whose entanglement monogamy score is \(\epsilon\). Therefore, the (two-dimensional) half-cone -like structure obtained in Theorem 1 contains 
state-points for all three-qubit pure states.

Theorem 1 therefore predicts an inverted two-dimensional conical shape for the state points in the \((\delta_C,{\cal E})\) plane, with one arm being 
vertical, and the other curved upwards. The curved line of 
the cone corresponds to generalized GHZ states, and the vertical line is the GGM axis.
Precisely, the 
curved line can be represented by the equation
\begin{equation}
{\cal E}=\frac{1}{2}(1-\sqrt{1-\delta_C}).
\end{equation}
The vertical line is, of course, represented by the equation \(\delta_C=0\). 
The tangent to the curved line at the tip of the cone (i.e. at the point where both GGM  and entanglement monogamy score vanish)
makes a nonzero angle with the axis of entanglement monogamy score. The tangent of this nonzero angle is \(\frac{1}{4}\).

In analogy with the space-time light cone, the tangent to the curved line and the GGM axis form the ``light cone'' 
of GGM and entanglement monogamy deficit. In this analogous ``relativity'', GGM acts as ``time''  and entanglement monogamy deficit acts as ``space'', 
and the ``velocity of light'' is \(\left.\frac{d\delta_C}{d{\cal E}}\right|_{\delta_C = 0}\), which is \(4\). Just like events cannot occur outside the 
space-time light cone,
multiparty quantum states cannot appear outside the light cone formed by entanglement monogamy deficit and GGM.

\subsubsection{Relating two multisite entanglement measures}
\label{lal-akash}

There are many entanglement measures that have been defined \cite{HHHH-RMP}, and it is important to try to find a  thread that connects them.  
The cone-like structure obtained in Theorem 1 implies a relation between the two multiparty entanglement measures. Precisely, it states that for three-qubit pure states, the GGM and the entanglement monogamy score (3-tangle)
are restricted by the relation
\begin{equation}
{\cal E}\geq \frac{1}{2}(1-\sqrt{1-\delta_C}).
\end{equation}

\subsubsection{Quantum discord as the quantum correlation measure in quantum monogamy score}

We will now prove a result for discord monogamy score, parallel to the one obtained in Sec. \ref{dos-ta-beyallis-atKGMarg}. 
The situation here is richer than that for entanglement monogamy score, and 
the complete picture is presented in two parts in the two following theorems.

The reason for the choice of quantum discord as the quantum correlation 
measure in the quantum monogamy score is two-fold: 
\begin{itemize}
\item[(i)] Quantum discord is a measure of quantum correlation
defined from a perspective that is independent of the 
entanglement-separability paradigm, the latter being 
the usual one for defining measures of quantum correlation (as for example, concurrence). 

\item[(ii)] Although quantum discord does not as yet have a closed 
form for arbitrary quantum states, it is possible to numerically calculate its 
values in an efficient manner.
\end{itemize}

Let the discord monogamy score of \(|\psi\rangle\) be denoted as  $\delta_{D}$,   and let 
$\delta_{D}^{GG}$ denote the same for the generalized GHZ state \(|GG(\alpha)\rangle\). \\
\noindent \textbf{Theorem 2a:} Consider the pure three-qubit state \(|\psi\rangle\) whose discord monogamy score is the
same as that of the generalized GHZ state \(|GG(\alpha)\rangle\). Then the genuine multipartite entanglement measure (GGM) will obey the ordering
\begin{equation}
{\cal E} \geq {\cal E}^{GG},
\end{equation}
with the nodal observer in the discord monogamy score being the one which contributes the maximal Schmidt coefficient in GGM.\\

\noindent \texttt{Proof:} 
 For an arbitrary three-qubit pure state $|\psi\rangle$, 
suppose that the quantum discords of the local density matrices $\rho_{AB}$ and $\rho_{AC}$ of   $|\psi\rangle$
are respectively $D_{AB}$ and $D_{AC}$. Further, let the entanglement of formations \cite{Eof} of $\rho_{AB}$ and $\rho_{AC}$ be respectively denoted 
by $E_{AB}$ and $E_{AC}$. 
The quantum discord of the pure state $|\psi\rangle$ in the bipartition of $A:BC$ is  the von Neumann entropy of the local density matrix $\rho_A$:
 we denote it by $S_A$ \cite{discord1, discord2}.
Now we use the Koashi-Winter relation \cite{Koashi} between quantum discord and entanglement of formation for the pure three-qubit state $|\psi\rangle$:
\begin{equation}
\label{eibar-Jasidih-elo}
 D_{AB} = E_{AC} - S_{A|B},
\end{equation}
where $S_{A|B}$ denotes the conditional entropy, and is defined as  $S_{A|B} = S(\rho_{AB}) - S(\rho_B)$. A similar equality holds for $D_{AC}$. 

Suppose now that the maximum in the GGM of \(|\psi\rangle\) is attained in \(\lambda_A^2\) (see Eq. (\ref{ekhono-Mughalsarai-aseni})). 
Applying the two relations obtained, respectively for \(D_{AB}\) and \(D_{AC}\), we find that
the monogamy score for quantum discord for an arbitrary three-qubit state, with \(A\) as the nodal observer,  as
\begin{equation}
 \label{asol_discord}
\delta_{D} = S_{A} - E_{AB} - E_{AC}.
\end{equation}
On the other hand, the monogamy score for the generalized GHZ state (for any observer as the nodal observer), obtained by using above equations in the specific case (of \(|GG(\alpha)\rangle\)), is given by 
\begin{equation}
 \label{asol_discord_GG}
\delta_{D}^{GG} = S_{A}^{GG}, 
\end{equation}
where $S_{A}^{GG}$ denotes the von Neumann entropy of a single-party local density matrix of the generalized GHZ state.
This is because the two-particle entanglements vanish for the generalized GHZ states. 
Consider now a generalized GHZ state whose discord monogamy score is the same as that of an arbitrary pure three-qubit state, i.e.,  
\begin{equation}
\delta_D = \delta_D^{GG}. 
\end{equation}
 This leads to 
\begin{eqnarray}
\label{keypoint}
S_A - E_{AB} - E_{AC} = S_{A}^{GG} ,\nonumber \\
\implies S_A - S_{A}^{GG} \geq 0, 
\end{eqnarray}
since the sum, $E_{AB} + E_{AC} $,  of the entanglements of formation  is always positive. 

Now suppose, if possible, that the GGMs for the arbitrary state and the generalized GHZ state does not obey the proposed ordering, i.e. suppose ${\cal E}^{GG} > {\cal E}$. This implies that 
$\lambda^2_A  < \alpha^2$.
This, along with the fact that \(\lambda_A^2\) (\(\alpha^2\)) is the \emph{maximal} eigenvalue of \(\rho_A^\psi\) (\(\rho_A^\alpha\)), leads to $S_A - S_{A}^{GG} < 0$, contradicting 
the inequality in (\ref{keypoint}).
 Hence the theorem. \hfill $\blacksquare$

The above theorem brings all pure three-qubit states that have a positive discord monogamy score into a half-cone -like structure in the space of \(\delta_D\) versus GGM.
Quantum discord can however be polygamous for three-qubit states \cite{monogamy-discord}, and hence there are states with a negative discord monogamy score \(\delta_D\). The generalized GHZ states, on the other hand,
always have a positive \(\delta_D\). The following theorem brings all pure three-qubit states with a negative discord monogamy score into a complementary half-cone -like structure by using the mirror image, with respect to the \(\delta_D=0\) axis,
 of the line 
created by the generalized GHZ states on the \((\delta_D,{\cal E})\) plane. \\
\textbf{Theorem 2b:} Consider the pure three-qubit state \(|\psi\rangle\) whose discord monogamy score is the negative 
of that of the generalized GHZ state \(|GG(\alpha)\rangle\). Then the genuine multipartite entanglement measure (GGM) will obey the ordering
\begin{equation}
{\cal E} \geq {\cal E}^{GG},
\end{equation}
with the nodal observer in the discord monogamy score being the one which contributes the maximal Schmidt coefficient in GGM.
\\

\noindent \texttt{Proof:} Let us first assume that 
the maximum in the GGM of \(|\psi\rangle\) is attained in \(\lambda_A^2\) (see Eq. (\ref{ekhono-Mughalsarai-aseni})). 
By using the Eq. (\ref{asol_discord}) in the enunciation, \(\delta_D = -\delta_D^{GG}\), with \(A\) as the nodal observer, we have
\begin{equation}
 S_A + S_{A}^{GG}  = E_{AB} + E_{AC}.
\end{equation}
But entanglements of formation of a bipartite state is no greater than either of the local von Neumann entropies \cite{Eof}, so that
\begin{equation}
 E_{AB} + E_{AC}  \leq 2 S_A,
 \end{equation}
whereby 
\begin{equation}
\label{aaj-Praci-aar-or-ma-amader-songe-school-thheke-jabe}
S_A - S_{A}^{GG} \geq 0.
\end{equation}
Assuming, if possible, that  ${\cal E}^{GG} > {\cal E}$, we obtain  $S_A - S_{A}^{GG}<0$,  contradicting the relation in 
(\ref{aaj-Praci-aar-or-ma-amader-songe-school-thheke-jabe}). 
Hence the theorem.
\hfill $\blacksquare$

So, while the state points in the \((\delta_C,{\cal E})\) plane are bounded by the curve for the generalized GHZ states and the \(\delta_C=0\) line, the same
in the \((\delta_D,{\cal E})\) plane are bounded by the curves for the generalized GHZ states and the mirror images of those curves with respect to the \(\delta_D=0\) line.
Precisely, they form an inverted two-dimensional conical shape, with the tip of the cone being at the point where
the GGM and the discord monogamy score are both zero,
and
the bounding 
curves are represented respectively by the equations
\begin{equation}
\delta_D=\pm({\cal E}\log_2{\cal E}+(1-{\cal E})\log_2(1-{\cal E})).
\end{equation}
The tangent to the boundary of the cone at the origin (i.e. at the point where both GGM  and discord monogamy score vanish)
is the discord monogamy score axis (\({\cal E}=0\)).

Continuing the analogy with the space-time light cone, we again have a ``light cone''-like structure between
the GGM and discord monogamy deficit. In this analogous ``relativity'', GGM acts as ``time''  and discord monogamy deficit acts as ``space'', 
and the ``velocity of light'' is \(\left.\frac{d\delta_D}{d{\cal E}}\right|_{\delta_D = 0}\), which is infinite. 


\subsection{Entanglement monogamy score and GGM}
\label{poune-baro-at-4-Ishan-Mondol-2}


In this subsection, 
we numerically simulate all pure states of three qubits, to obtain a scatter diagram to see the inter-relation between the entanglement monogamy score (3-tangle) 
and the generalized geometric measure, obtained in Theorem 1. 
The scatter plot is given in Fig. 1. Since 
entanglement monogamy score is always non-negative (because the concurrence squared satisfies 
the monogamy relation \cite{Wootters, Koashi, monogamyN}), and as 
the GGM is always non-negative, all state points are in the first quadrant of  
 the (entanglement monogamy score, GGM) plane.
\begin{figure}%
\begin{center}
\includegraphics[width=0.6\columnwidth]{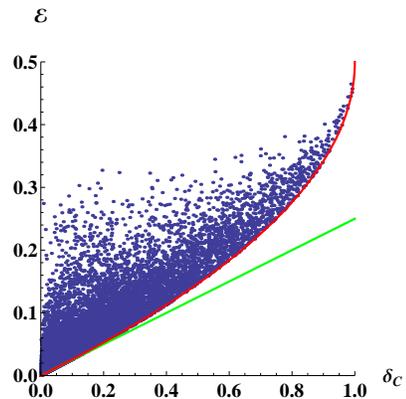}%
\label{fig-chhobi-asol-maney-jetate-sob-state-achhe}
\caption{(Color online.)
Multiparty entanglement vs. quantum monogamy score. Here, quantum monogamy score is identified with the entanglement monogamy score, and multiparty 
entanglement is quantified by generalized geometric measure, which is a genuine multisite entanglement. The units of the axes are chosen as follows. 
The vertical axis is in units for which the GGM of the 
GHZ state (generalized GHZ state with \(\alpha=\frac{1}{\sqrt{2}}\)) is \(\frac{1}{2}\), while the horizontal axis is such that the concurrence of the singlet state
is unity. For a randomly chosen pure state of three qubits, its entanglement monogamy score is taken as the abscissa and its GGM as the ordinate, and 
the pair is then plotted respectively on the horizontal and vertical 
axes of the figure. We randomly choose \(2.5 \times 10^4\) points. The envelope of the points thus plotted forms a two-dimensional conical shape, with
 one axis coinciding with the vertical axis (GGM axis, i.e., 
entanglement monogamy score \(= 0\)), and the other being formed out of generalized GHZ states. The tangent to the curved boundary at the origin of the 
two axis forms a nonzero angle with the horizontal axis 
(entanglement monogamy score axis, i.e., GGM \(= 0\)).}
\end{center}
\end{figure} 

Fig. 1 clearly  confirms  that for a given value of the entanglement monogamy score, 
all three-qubit pure states have their genuine multiparty entanglement, as 
quantified by the GGM, higher than a certain value, and that state points are constrained to lie within a cone formed by the lines 
\begin{equation}
{\cal E}=\frac{1}{2}(1-\sqrt{1-\delta_C}), \quad \delta_C=0.
\end{equation}

\subsection{Discord monogamy score and GGM}
\label{poune-baro-at-4-Ishan-Mondol-3}

\begin{figure}%
\begin{center}
\includegraphics[width=0.6\columnwidth]{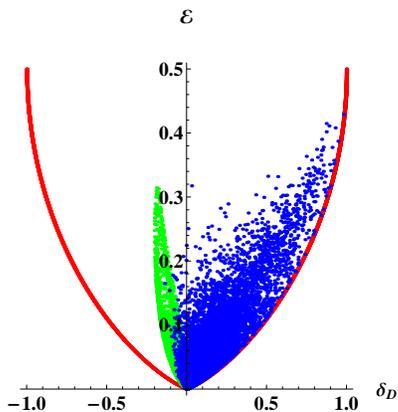}
\label{fig-chhobi-eita-discord-tangle-er-songey-gigiem}
\caption{(Color.) Light cone of discord monogamy score and GGM. The notation is just like in Fig. 1,
 except that the horizontal axis now depicts discord monogamy score, 
and is measured in bits. We randomly generate \(5 \times 10^4\) points. The envelope this time is formed of 
two curved lines, of which the right one is again (i.e., like in Fig. 1) 
formed out of generalized GHZ states, while the one on the left is the mirror image of the one on the right with respect 
to the GGM axis. The tangent to the curves at the origin of the two axes in this case coincides with the GGM \(= 0\) line. 
While the blue points represent randomly chosen points from the GHZ-class, the green ones represent those from the W-class.
}
\end{center}
\end{figure}


In this subsection, 
numerical simulation of all three-qubit pure states will be used to obtain a scatter diagram 
to look at the relation between the GGM and the quantum monogamy score, with quantum discord being used as the measure of quantum correlation in the monogamy score. 
This will give us a pictorial view of the results presented in Theorems 2a and 2b. 
In Fig. 2, we plot the discord monogamy score against GGM, for pure 
three-qubit states. The quantum state points in the figure are 
randomly chosen from the GHZ-class states \cite{dur-vidal-cirac} and  W-class states \cite{Wstate,dur-vidal-cirac}. 
As expected from Theorems 2a and 2b, they again form an inverted two-dimensional conical shape, with the tip of the cone being at the point where
the GGM and the discord monogamy score are both zero, with the bounding curve being represented 
 by the equations
\begin{equation}
\delta_D=\pm({\cal E}\log_2{\cal E}+(1-{\cal E})\log_2(1-{\cal E})).
\end{equation}
It is interesting to note that the obtained cone is independent of the chosen nodal observer.

\subsection{Entanglement versus Discord}

Despite the similarities, 
there are 
interesting differences between the conical shapes 
obtained for entanglement monogamy score and discord monogamy score: 
\begin{itemize}
\item[(1)] There are two curved lines that form the cone on the plane of discord monogamy score and GGM, while that in the plane of entanglement monogamy score 
and GGM is formed out of one curved line and a straight (vertical) line. This is a consequence of the fact 
that while the 3-tangle respects monogamy (entanglement monogamy score is always positive \cite{Wootters, Koashi, monogamyN}), quantum discord does not 
(discord monogamy score can be both negative and positive \cite{monogamy-discord}).  
\item[(2)] The tangent to the curved lines at the tip of the cone is the discord monogamy score axis. 
In other words, \(\left.\frac{d\delta_D}{d{\cal E}}\right|_{\delta_D=0} = \infty\), while \(\left.\frac{d\delta_C}{d{\cal E}}\right|_{\delta_C=0} = 4\).
\end{itemize}

\subsection{Relating the two main paradigms for defining quantum correlations in the multisite domain}
\label{holde-sobuj-orangotang}

It is interesting to note here that discord monogamy score is closely related to the multiparty information-theoretic quantum correlation 
measure called dissension \cite{Arun-disso}. 
%
%
%
Dissension is defined as the difference between the three-variable mutual information with conditional 
entropies involving no measurements, and with measurements on various subsystems
\cite{Arun-disso}. It is apparent that there can be different kinds of dissensions depending on the type of measurement involved. 
The definition we choose here is presented below.\\

\noindent {\bf Definition I:} The quantum mutual information of a three-party quantum state $\rho_{ABC}$ is defined as
$$I(\rho_{ABC})=I(\rho_{AB})-I(\rho_{A:B|C}),$$
where $I(\rho_{AB})$ is defined in Eq. (\ref{qmi1}), and the unmeasured conditional quantum mutual information is defined as
\begin{equation}
I(\rho_{A:B|C})=\tilde S(\rho_{A|C})-\tilde S(\rho_{A|BC}),
\label{eq:qmidis}
\end{equation}
with
\begin{eqnarray}
\tilde S(\rho_{A|C})&=&S(\rho_{AC})-S(\rho_{C}),\nonumber\\
\tilde S(\rho_{A|BC})&=&S(\rho_{ABC})-S(\rho_{BC}).\nonumber
\end{eqnarray}
In the same spirit as in Ref. \cite{discord1,discord2} (cf. \cite{Arun-disso}), we can define the three-party classical mutual information for 
the quantum state $\rho_{ABC}$ as
$$J(\rho_{ABC})=J(\rho_{AB})-J(\rho_{A:B|C}),$$
where $J(\rho_{AB})$ is defined in Eq. (\ref{cmi1}), and the measured conditional quantum mutual information is defined as 
\begin{equation}
J(\rho_{A:B|C})=S(\rho_{A|C})-S(\rho_{A|BC}),
\label{eq:cmidis}
\end{equation}
where the quantum conditional entropies are defined in Eqns. (\ref{qce}).
Dissension can now be defined as
\begin{equation}
D(A:B:C)=I(\rho_{ABC})-J(\rho_{ABC}).
\label{eq:dissension}
\end{equation}

Note that that in Ref. \cite{Arun-disso}, there are two kinds of dissension used which involves single-particle and two-particle measurements separately. 
In contrast, the variety of dissension defined here 
involves both single-particle and two-particle measurements on $B,\, C,$ and $BC$ respectively.\\
\noindent {\bf Proposition I:} For a tripartite quantum state $\rho_{ABC}$, the dissension 
$D(A:B:C)=D(\rho_{AB})+D(\rho_{AC})-D(\rho_{A:BC})$, the negative of discord monogamy score of $\rho_{ABC}$.\\

\noindent \texttt{Proof:} We have 
\begin{eqnarray}
D(A:B:C)&=&I(\rho_{ABC})-J(\rho_{ABC})\nonumber\\
&=&I(\rho_{AB})-J(\rho_{AB})\nonumber\\
& &\hspace{1em} -\left[I(\rho_{A:B|C})-J(\rho_{A:B|C})\right]\nonumber\\
&=&D(\rho_{AB})-\left[I(\rho_{A:B|C})-J(\rho_{A:B|C})\right].\nonumber
\end{eqnarray}
Using the definitions of $I(\rho_{A:B|C})$ and $J(\rho_{A:B|C})$, given in Eqns. (\ref{eq:qmidis}) and (\ref{eq:cmidis}) respectively, we have
$$I(\rho_{A:B|C})-J(\rho_{A:B|C})=D(\rho_{A:BC})-D(\rho_{AC}).$$
Hence the proof. \hfill \(\blacksquare\)

There is an ongoing effort to connect the quantum correlation measures defined in the two main paradigms, viz. the entanglement-separability and the 
information-theoretic ones. See Refs. \cite{Koashi, discord-surge}. 
The theorems 2a and 2b obtained in this paper tries to find similar connections in the multipartite domain. Precisely, we find that 
for pure three-qubit states, the generalized geometric measure (a genuine \emph{multisite entanglement} measure) and the discord monogamy score 
(a \emph{multisite information-theoretic} quantum correlation measure, via proposition I), are constrained by 
\begin{equation}
\label{sobuj-prithhibi}
\delta_D \leq \left|{\cal E}\log_2{\cal E}+(1-{\cal E})\log_2(1-{\cal E})\right|.
\end{equation}


\section{Discussion}
\label{sec-ekhane-conclu}

The monogamy relation is an important tool to decipher the structure of the space of quantum correlation measures. 
There are measures that satisfy and those that violate this relation. 
We have shown that given a certain  amount of violation or satisfaction of the monogamy relation, the allowed range of the 
genuine multisite entanglement content of the corresponding pure three-qubit state is distinctly restricted. The quantum states of the corresponding 
system is thereby restricted to remain within an envelope in the plane formed by the quantum monogamy score and genuine multiparty entanglement. 
We have used the generalized geometric measure for quantifying genuine multiparty entanglement. Quantum monogamy score, on the other hand, is defined by using 
two measures of bipartite quantum correlation -- first by using the square of the  concurrence, and then by quantum discord.

The relations thus obtained between quantum monogamy scores and genuine multiparty entanglement, currently only for pure three-qubit states, is akin to that between space and time in the theory of relativity.

The quantum monogamy score, defined by using concurrence squared, has been proposed as a measure of multipartite entanglement. Also, quantum monogamy score, defined by 
using quantum discord, is very similar to a proposed information-theoretic measure of multiparty quantum correlation, called dissension. Therefore, the results obtained 
imply that 
just like space and time are intertwined in the  theory of relativity and thereby future event points are constrained to lie within the future cone,
the apparently different multiparty quantum correlation concepts can be intertwined, thereby constraining the multipartite quantum state space within a  conical structure.
 We hope that a better picture will emerge in future that will unify the multiparticle entanglement measures and multiparticle correlations. We believe that our results 
form a first step in this direction.

\acknowledgments
We acknowledge computations performed at the cluster computing facility in HRI (\url{http://cluster.hri.res.in/}).


\begin{thebibliography}{999}

\bibitem{ekhane-NC} M.A. Nielsen and I.L. Chuang, \emph{Quantum Computation and Quantum Information} 
(Cambridge University Press, Cambridge, 2000).

\bibitem{HHHH-RMP} R. Horodecki,
 P. Horodecki,  M. Horodecki, and K. Horodecki,
 Rev. Mod. Phys. \textbf{81}, 865 (2009).


\bibitem{comm-review} For a recent review, see e.g., 
A. Sen(De) and U. Sen, Physics News \textbf{40}, 17 (2010) (arXiv:1105.2412 [quant-ph]). 



\bibitem{dc} C.H. Bennett and S.J. Wiesner, Phys. Rev. Lett \textbf{69}, 2881 (1992).

\bibitem{tele} C.H. Bennett, G. Brassard, C. Cr{\'e}peau, R. Jozsa, A. Peres, and  W.K. Wootters, Phys. Rev. Lett. \textbf{70}, 1895 (1993).

\bibitem{Ekert91} A.K. Ekert,  Phys. Rev. Lett. {\bf 67}, 661 (1991).


\bibitem{remote} A.K. Pati, Phys. Rev. A \textbf{63}, 014302 (2000); C.H. Bennett, D.P. DiVincenzo, P.W. Shor, J.A. Smolin, B.M. Terhal, and W.K. Wootters,
Phys. Rev. Lett. \textbf{87}, 077902 (2001).

\bibitem{Briegel} H.J. Briegel, D.E. Browne, W. D{\"u}r, R. Raussendorf, and M. Van den Nest, Nat. Phys. \textbf{5}, 19 (2009).



\bibitem{nlwe} C.H. Bennett, D.P. DiVincenzo, C.A. Fuchs, T. Mor, E. Rains, P.W. Shor, J.A. Smolin, and W.K. Wootters, Phys. Rev. A \textbf{59}, 1070 (1999);
C.H. Bennett, D.P. DiVincenzo, T. Mor, P.W. Shor, J.A. Smolin, and B.M. Terhal,
Phys. Rev. Lett. \textbf{82}, 5385 (1999); D.P. DiVincenzo, T. Mor, P.W. Shor, J.A. Smolin, and
B.M. Terhal, Comm. Math. Phys. \textbf{238}, 379 (2003).

\bibitem{local-indis-other} A. Peres and W.K. Wootters, Phys. Rev. Lett. \textbf{66}, 1119 (1991);
J. Walgate, A.J. Short, L. Hardy, and V. Vedral, \emph{ibid.} \textbf{85}, 4972 (2000); 
S. Virmani, M.F. Sacchi, M.B. Plenio, and D. Markham, Phys. Lett. A \textbf{288}, 62 (2001); 
Y.-X. Chen and D. Yang, Phys. Rev. A \textbf{64}, 064303 (2001); \textbf{65}, 022320 (2002); 
J. Walgate and L. Hardy, Phys. Rev. Lett. \textbf{89}, 147901 (2002);
M. Horodecki, A. Sen(De), U. Sen, and K. Horodecki, \emph{ibid.} \textbf{90}, 047902 (2003); 
W.K. Wootters, IJQI, {\bf 4}, 219 (2006).

\bibitem{KnillLaflamme}E. Knill and R. Laflamme, Phys. Rev. Lett. \textbf{81}, 5672 (1998).

\bibitem{Animesh} A. Datta, A. Shaji, and C. M. Caves, Phys. Rev. Lett. \textbf{100}, 050502 (2008).

\bibitem{others} S.L. Braunstein, C.M. Caves, R. Jozsa, N. Linden, S. Popescu, and R. Schack, Phys. Rev. Lett. \textbf{83}, 1054 (1999); 
D.A. Meyer, \emph{ibid.} \textbf{85}, 2014 (2000);
S.L. Braunstein and A.K. Pati, Quant. Inf. Comp. \textbf{2}, 399 (2002);
A. Datta, S.T. Flammia, and C.M. Caves, Phys. Rev. A \textbf{72},
042316 (2005);
A. Datta and G. Vidal, \emph{ibid.} \textbf{75}, 042310 (2007); 
B.P. Lanyon, M. Barbieri, M.P. Almeida, and A.G. White , Phys. Rev. Lett. \textbf{101}, 200501 (2008).




\bibitem{discord2} L. Henderson, and V. Vedral, J. Phys. A \textbf{34}, 6899 (2001).

\bibitem{discord1} H. Ollivier and W. H. Zurek, Phys. Rev. Lett. \textbf{88}, 017901
(2001).

\bibitem{workdeficit}J. Oppenheim, M. Horodecki, P. Horodecki, and R. Horodecki,
Phys. Rev. Lett. \textbf{89}, 180402 (2002); M. Horodecki, K. Horodecki, P. Horodecki, 
R. Horodecki, J. Oppenheim, A. Sen(De), and U. Sen, \emph{ibid.} \textbf{90}, 100402 (2003); 
M. Horodecki, P. Horodecki, R. Horodecki, J. Oppenheim, A. Sen(De), U. Sen, and B. Synak-Radtke,
Phys. Rev. A \textbf{71}, 062307 (2005). 


\bibitem{Modi} K. Modi, T. Paterek, W. Son, V. Vedral, and M. Williamson,
Phys. Rev. Lett. \textbf{104}, 080501 (2010).

\bibitem{Arun-disso} I. Chakrabarty, P. Agrawal, and A.K. Pati,	EPJD {\bf 65}, 605 (2011).

\bibitem{discord-multi} M. Okrasa and Z. Walczak, 
Europhys. Lett., \textbf{96}, 60003 (2011);
C.C. Rulli and M.S. Sarandy, 
Phys. Rev. A \textbf{84}, 042109 (2011);
Z.-H. Ma and Z.-H. Chen, arXiv:1108.4323 [quant-ph].


\bibitem{bakisob}  C. A. Rodr{\'i}guez-Rosario, K. Modi, A.-M. Kuah, A. Shaji, and 
E.C.G. Sudarshan, J. Phys. A: Math. Theor. \textbf{41}, 205301 (2008); 
M. Piani, P. Horodecki, and R. Horodecki, Phys. Rev. Lett. \textbf{100}, 090502 (2008); 
A. Shabani and D.A. Lidar, Phys. Rev. Lett. \textbf{102}, 100402
(2009); S. Boixo, L. Aolita, D. Cavalcanti, K. Modi, and A. Winter, 
Int. J. Quant. Inf. \textbf{9}, 1643 (2011);
L. Wang, J.-H. Huang, J.P. Dowling, and S.-Y. Zhu, arXiv:1106.5097 [quant-ph], and references therein.

\bibitem{Wootters} V. Coffman, J. Kundu, and W.K. Wootters, Phys. Rev. A \textbf{61}, 052306 (2000).

\bibitem{Koashi}M. Koashi and A. Winter, Phys. Rev. A \textbf{69}, 022309 (2004).


\bibitem{monogamyN} See also 
T.J. Osborne and F. Verstraete, Phys. Rev. Lett. \textbf{96}, 220503 (2006);
G. Adesso, A. Serafini, and F. Illuminati, Phys. Rev. A \textbf{73}, 032345 (2006);
T. Hiroshima, G. Adesso, and F. Illuminati, Phys. Rev. Lett. \textbf{98}, 050503 (2007); 
M. Seevinck, Phys. Rev. A \textbf{76}, 012106 (2007); 
S. Lee and J. Park, Phys. Rev. A \textbf{79}, 054309 (2009);
 A. Kay, D. Kaszlikowski, and R. Ramanathan, Phys. Rev. Lett. \textbf{103}, 050501 (2009);
 M. Seevinck, QIP {\bf 9}, 273 (2010); 
 M.-J. Zhao, S.-M. Fei, and Z.-X. Wang, Int. J. Quant. Inform. {\bf 8}, 905 (2010); 
 M. Hayashi and L. Chen, Phys. Rev. A \textbf{84}, 012325 (2011);
 J.S. Kim and B.C. Sanders, J. Phys. A: Math. Theor. {\bf 44}, 295303 (2011);
F.F. Fanchini, M.F. Cornelio, M.C. de Oliveira, and A.O. Caldeira, Phys. Rev. A \textbf{84}, 012313 (2011);
 G.L. Giorgi, 
Phys. Rev. A \textbf{84}, 054301 (2011), and references therein.

\bibitem{monogamy-discord} R. Prabhu, A.K. Pati, A. Sen(De), and U Sen, 
Phys. Rev. A \textbf{R85}, 040102 (2012). 

\bibitem{monogamy-iacc}A. Sen(De) and U. Sen, Phys. Rev. A \textbf{85}, 052103 (2012).



\bibitem{Eof} C.H. Bennett, H.J. Bernstein, S. Popescu, and B. Schumacher, Phys. Rev. A {\bf 53}, 2046 (1996); 
C.H. Bennett, D.P. DiVincenzo, J.A. Smolin, and W.K. Wootters, Phys. Rev. A {\bf 54}, 3824 (1996).

\bibitem{concurrence} S. Hill and W.K. Wootters, Phys. Rev. Lett. {\bf 78}, 5022 (1997); W.K. Wootters, Phys. Rev. Lett. {\bf 80}, 2245 (1998).





\bibitem{discord-surge} H.S. Dhar, R. Ghosh, A. Sen(De), and U. Sen, EPL  \textbf{98}, 30013 (2012); A. Sen(De), U. Sen, and M. Lewenstein,
Phys. Rev. A \textbf{72}, 052319 (2005).

\bibitem{logneg} G. Vidal and R.F. Werner, Phys. Rev. A \textbf{65}, 032314 (2002).

\bibitem{GGM} A. Sen(De) and U. Sen, Phys. Rev. A {\bf 81}, 012308 (2010); 
A. Sen(De) and U. Sen, arXiv:1002.1253 [quant-ph];
R. Prabhu, S. Pradhan, A. Sen(De), and U. Sen, Phys. Rev. A {\bf 84}, 042334 (2011).

\bibitem{GHZ}D.M. Greenberger, M.A. Horne, and A. Zeilinger, in \emph{Bell's Theorem, Quantum Theory, and Conceptions of
the Universe}, ed. M. Kafatos (Kluwer Academic, Dordrecht,  1989).

 
\bibitem{qmi}W.H. Zurek, in \emph{Quantum Optics, Experimental Gravitation
and Measurement Theory}, eds. P. Meystre and M.O. Scully (Plenum, New York, 1983);
S.M. Barnett and S.J.D. Phoenix, Phys. Rev. A \textbf{40}, 2404 (1989). 



\bibitem{Cerf} N.J. Cerf and C. Adami, Phys. Rev. Lett. \textbf{79}, 5194 (1997).


\bibitem{GROIS} B. Schumacher and M.A. Nielsen, Phys. Rev. A \textbf{54}, 2629 (1996); 
B. Groisman, S. Popescu, and A. Winter, Phys. Rev. A \textbf{72}, 032317 (2005).

\bibitem{GM} A. Shimony, Ann. N.Y. Acad. Sci. {\bf 755}, 675 (1995); H. Barnum and N. Linden, J. Phys. A {\bf 34}, 6787 (2001); 
T.-C. Wei and P.M. Goldbart, Phys. Rev. A {\bf 68}, 042307 (2003).

 



\bibitem{dur-vidal-cirac}  W. D{\"u}r, G. Vidal, and J.I. Cirac, Phys. Rev. A \textbf{62}, 062314 (2000).

\bibitem{Wstate} A. Zeilinger, M.A. Horne, and D.M. Greenberger, in \emph{Proceedings
of Squeezed States and Quantum Uncertainty}, eds. D. Han, Y.S. Kim, and W.W. Zachary, NASA Conf.
Publ. 3135 (1992).




\end{thebibliography}
\end{document}